%% file: ms.tex
\documentclass[apjl]{emulateapj}

\shorttitle{Cluster Radio Relic Discovered in Abell 2443}
\shortauthors{Cohen \& Clarke}

\begin{document}

\title{An Ultra-Steep Spectrum Radio Relic \\
in the Galaxy Cluster Abell 2443}

\author{A.~S.~Cohen \altaffilmark{1,2} and T.~E.~Clarke \altaffilmark{1}}
\altaffiltext{1}{Naval Research Laboratory, Code 7213, Washington, DC, 
20375 USA, aaron\_cohen@alum.mit.edu}
\altaffiltext{2}{Johns Hopkins University Applied Physics Laboratory
11100 Johns Hopkins Road, Laurel, MD 20723}


\begin{abstract}
We present newly discovered radio emission in the galaxy cluster Abell
2443 which is (1) diffuse, (2) extremely steep spectrum, (3) offset
from the cluster center, (4) of irregular morphology and (5) not
clearly associated with any of the galaxies within the cluster.  The
most likely explanation is that this emission is a cluster radio
relic, associated with a cluster merger.  We present deep observations
of Abell 2443 at multiple low frequencies (1425, 325 and 74~MHz) which
help characterize the spectrum and morphology of this relic. Based on
the curved spectral shape of the relic emission and the presence of
small scale structure, we suggest that this new source is likely a
member of the radio phoenix class of radio relics.

\end{abstract}


\section{Introduction}

A number of clusters of galaxies show extended synchrotron emission
not directly associated with the galaxies, but rather diffused in the
intra-cluster medium (ICM) \citep{feretti05}.  These radio sources
have low surface brightness, a variety of sizes up to $\sim$1 Mpc and
steep radio spectrum. They have been observationally classified as
halos and relics.  Halos are located at the cluster centers, show a
rather regular structure and little or negligible polarized emission.
Relics are at the cluster peripheries, are generally elongated in
shape and are often highly polarized. \citet{kempner04} further
classified radio relics into categories associated with extinct or
dying radio galaxies (radio phoenix and AGN relics) or those
associated more generally with the ICM (radio gischt). About 60
clusters are known to contain halos and or relics \citep{ferrari2009}.
The origin and evolution of this diffuse emission is still a matter of
debate, though the existence of powerful diffuse sources only in
clusters undergoing major mergers indicates that they are somehow
related to the merger process.  The detection and study of new objects
of this class is of great importance to establish correlations between
the radio source properties and the parent cluster properties, and to
discriminate among the proposed theoretical models of their origins
(see \citet{ferrari08} and references within).

\section{The Galaxy Cluster Abell 2443}

Abell 2443 is a rich cluster at an intermediate redshift of z = 0.108
\citep{struble99} with a relatively low X-ray luminosity of $L_x =
1.9\times10^{44} erg\,s^{-1}$ \citep{ebeling98}.  It contains a radio
source, catalogued as 4C+17.89.  A multi-color photometric study of
the system by \citet{wen07} identified 289 new cluster member
galaxies. Combining these with the 12 known member galaxies revealed a
north-west to south-east elongated galaxy distribution for the main
cluster connecting to a group in the south-east \citep[Figure
  6]{wen07} which they have identified with the galaxy cluster
ZwCl\ 2224.2+1651. There is an X-ray peak in the vicinity of the
southeast cluster but no redshift information is available. It is
unclear whether the two cluster systems are in the process of merging
or whether the images are just revealing the chance superposition of
two systems at different redshifts.

Comparing the Abell 2443 image in both the 1400~MHz NVSS \citep{condon98} and
74~MHz VLSS \citep{cohen07} indicated that Abell 2443 contains
diffuse steep-spectrum radio emission, but neither image provided the
resolution or sensitivity to confirm its nature, or distinguish it
from radio galaxy emission from cluster members.  We therefore
conducted a series of deeper, higher-resolution observations at
1425~MHz, 325~MHz and 74~MHz using the National Radio Astronomy
Observatory's Very Large Array radio telescope.

\section{Observations}

\subsection{Low-Resolution Observations}

\begin{figure}
\plotone{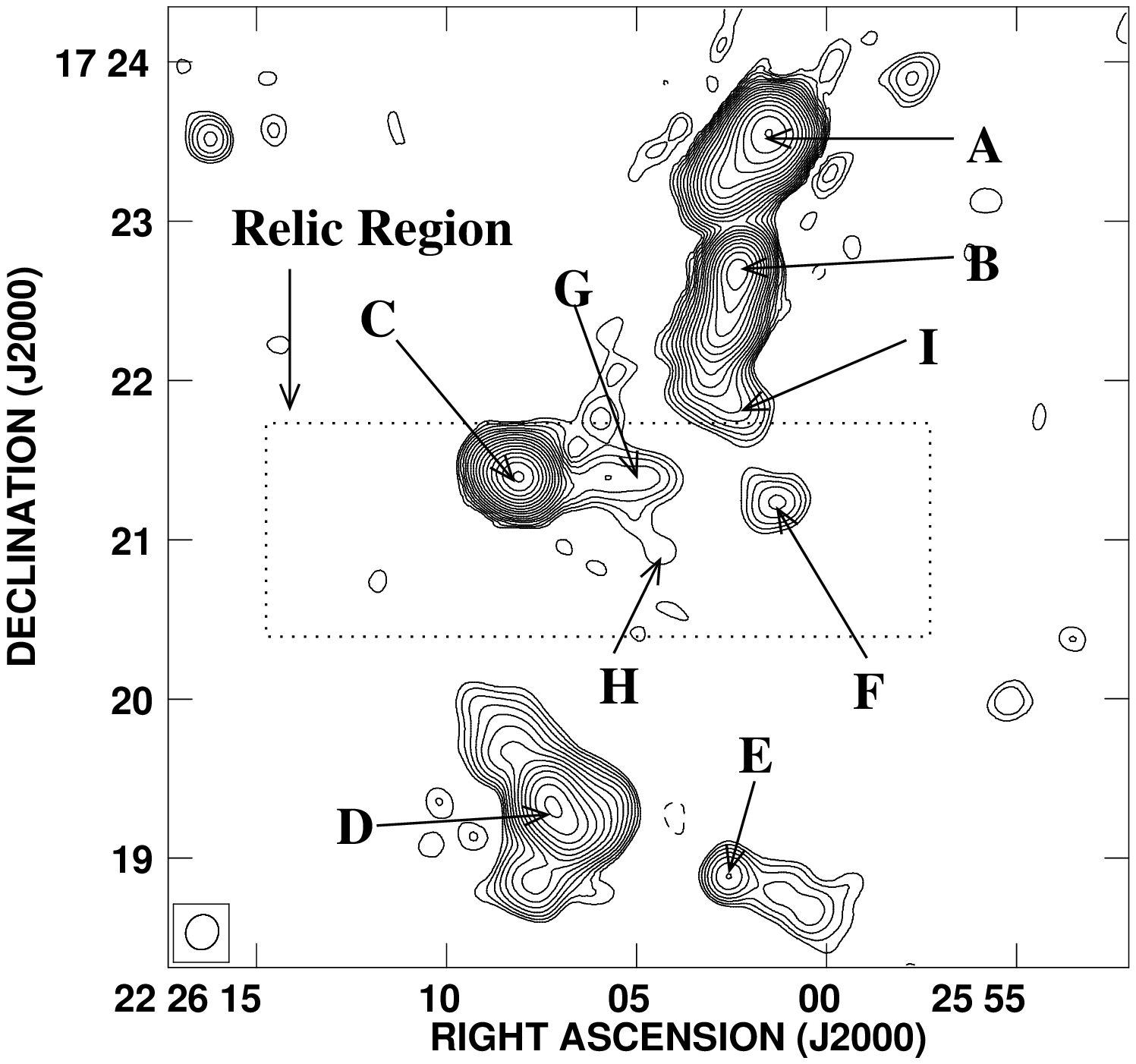}
\caption{Abell 2443 at 1425~MHz with VLA C-configuration.  
Contours begin at $\pm$0.148 mJy/beam ($3\times\sigma_{rms}$) and 
increase by multiples of $\sqrt{2}$.  The peak intensity is 38.4 mJy/beam.
\label{LC.fig}}
\end{figure}

\begin{figure}
\plotone{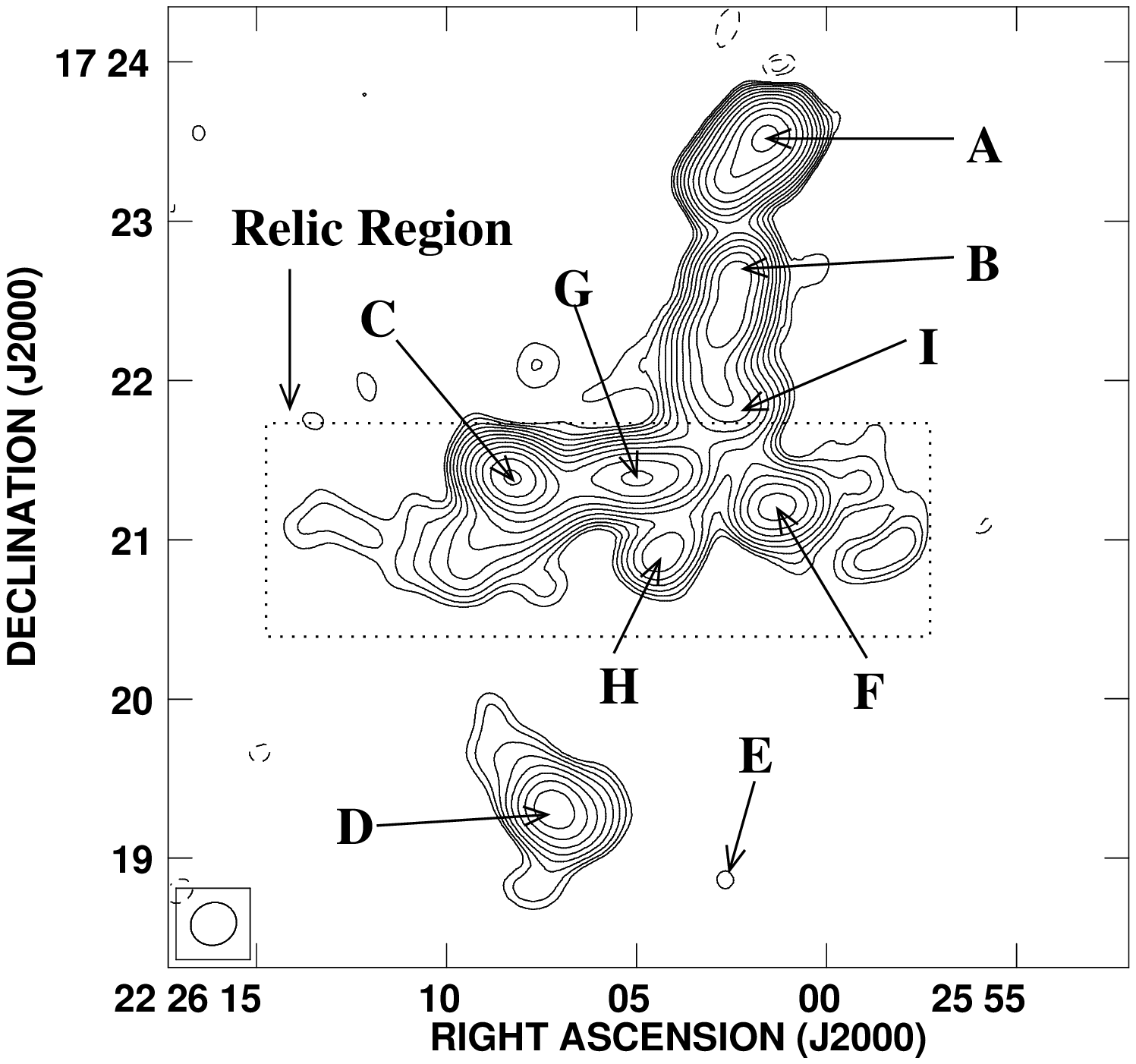}
\caption{Abell 2443 at 325~MHz with VLA B-configuration.
Contours begin at $\pm$2.41 mJy/beam ($3\times\sigma_{rms}$) and 
increase by multiples of $\sqrt{2}$.  The peak intensity is 125 mJy/beam.
\label{PB.fig}}
\end{figure}

\begin{figure}
\plotone{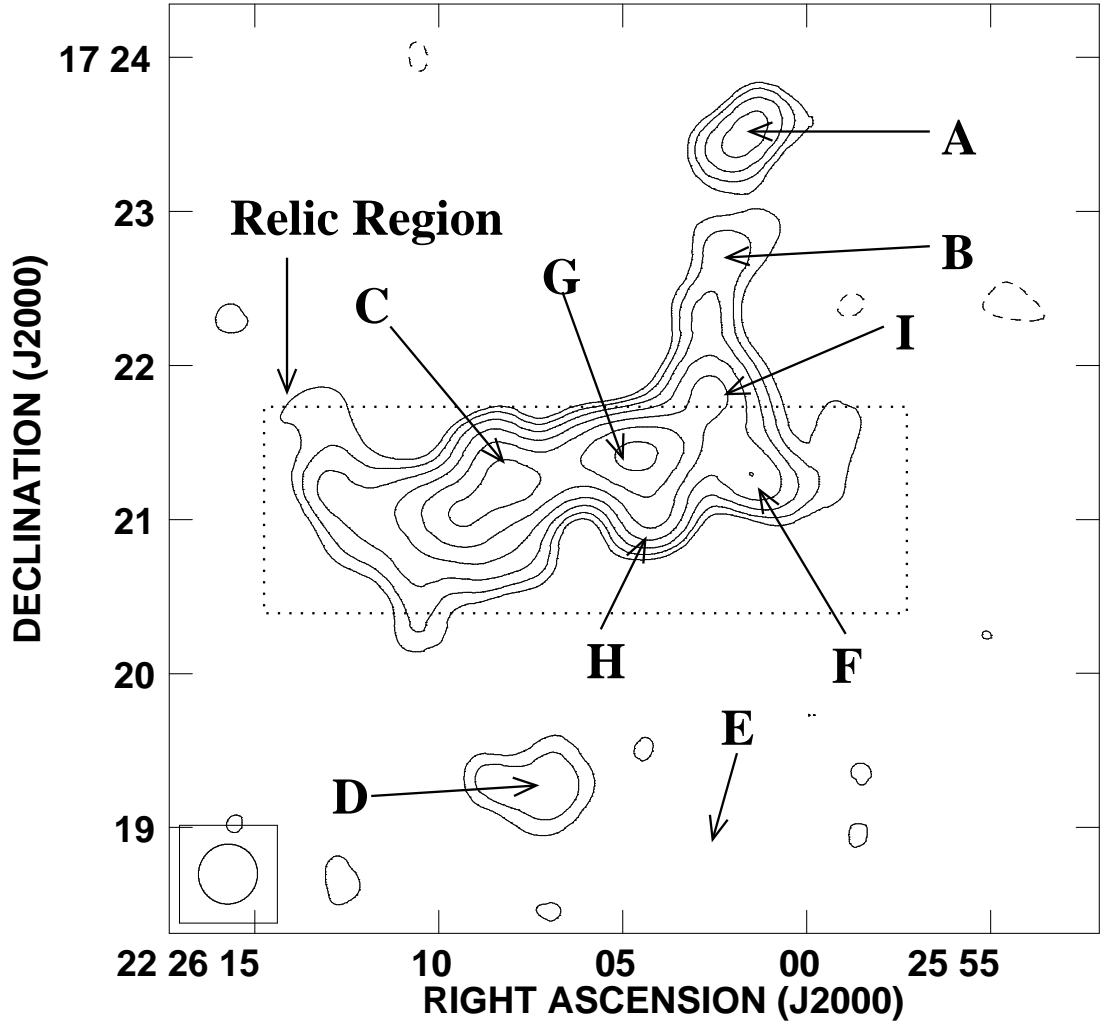}
\caption{Abell 2443 at 73.8~MHz with VLA A-configuration.
Contours begin at $\pm$88.2 mJy/beam ($3\times\sigma_{rms}$) and 
increase by multiples of $\sqrt{2}$.  The peak intensity is 778 mJy/beam.
\label{4A.fig}}
\end{figure}

Our observations were designed to map the radio emission in Abell 2443
at three low frequencies, (1425, 325 and 74~MHz).  By using different
VLA configurations (C, B and A, respectively), we obtained images of
roughly equivalent resolution at each of these three frequencies.  The resolution of between about
13$''$ and 23$''$ is optimal for surface-brightness sensitivity of the
diffuse emission \cite[see e.g.\ recent papers by][]{weeren,simona}.
The observational parameters are listed in
Table~\ref{observation.tab}, and the maps are shown in
Figures~\ref{LC.fig}, \ref{PB.fig} and \ref{4A.fig}, respectively.

Figures~\ref{LC.fig}, \ref{PB.fig} and \ref{4A.fig} are all shown at
the same scale and are of roughly the same resolution.  Together they
demonstrate that the radio morphology of Abell 2443 changes greatly
from 1425~MHz to 325~MHz and then to 74~MHz.  This is caused by
drastically different spectra among the various radio sources within
the cluster.  The sources seem to be split into two main groups.
First are the sources that appear to be head-tail radio galaxies.
These sources are A, B, C, D and E, which are best seen at 1425~MHz
(Figure~\ref{LC.fig}).  These sources appear as expected in the
325~MHz and 74~MHz maps (Figures \ref{PB.fig} and \ref{4A.fig},
respectively) assuming typical spectral indices for radio galaxies of
approximately $\alpha \approx -0.7$ ($S \propto \nu^\alpha$).  The
signal-to-noise ratio declines at lower frequencies because the map
noise increases greatly (Table~\ref{observation.tab}) due to the very high system
temperatures.  The second group of
sources include sources F, G, H, and I (see also Section 3.2) which
are best seen in the 325~MHz image (Figure~\ref{PB.fig}).  That they
are seen in the 325~MHz image with higher signal-to-noise ratio than
in the 1425~MHz image indicates a very steep spectrum.  In fact, a
point source would need a spectral index of $\alpha \leq -1.89$ to be
better detected in the 325~MHz image than in the 1425~MHz image.  Also
seen in the 325~MHz image are additional steep spectrum emission that 
extends the "global relic" both to the southeast, past source C, and 
to the west beyond source F, which is not seen
at all in the 1425~MHz image.  All the steep spectrum emission
combines to forms a roughly east-west ``arc'' throughout a region
indicated by the dotted rectangle in Figures \ref{LC.fig},
\ref{PB.fig} and \ref{4A.fig}.  In the 74~MHz image, this steep
spectrum arc, which was barely seen at 1425~MHz, is the most prominent
feature, with the flatter spectrum radio galaxies now relatively far
fainter. Hereafter we refer to the steep spectrum arc as the radio
relic.

\subsection{High-Resolution Observations}

\begin{figure}
\plotone{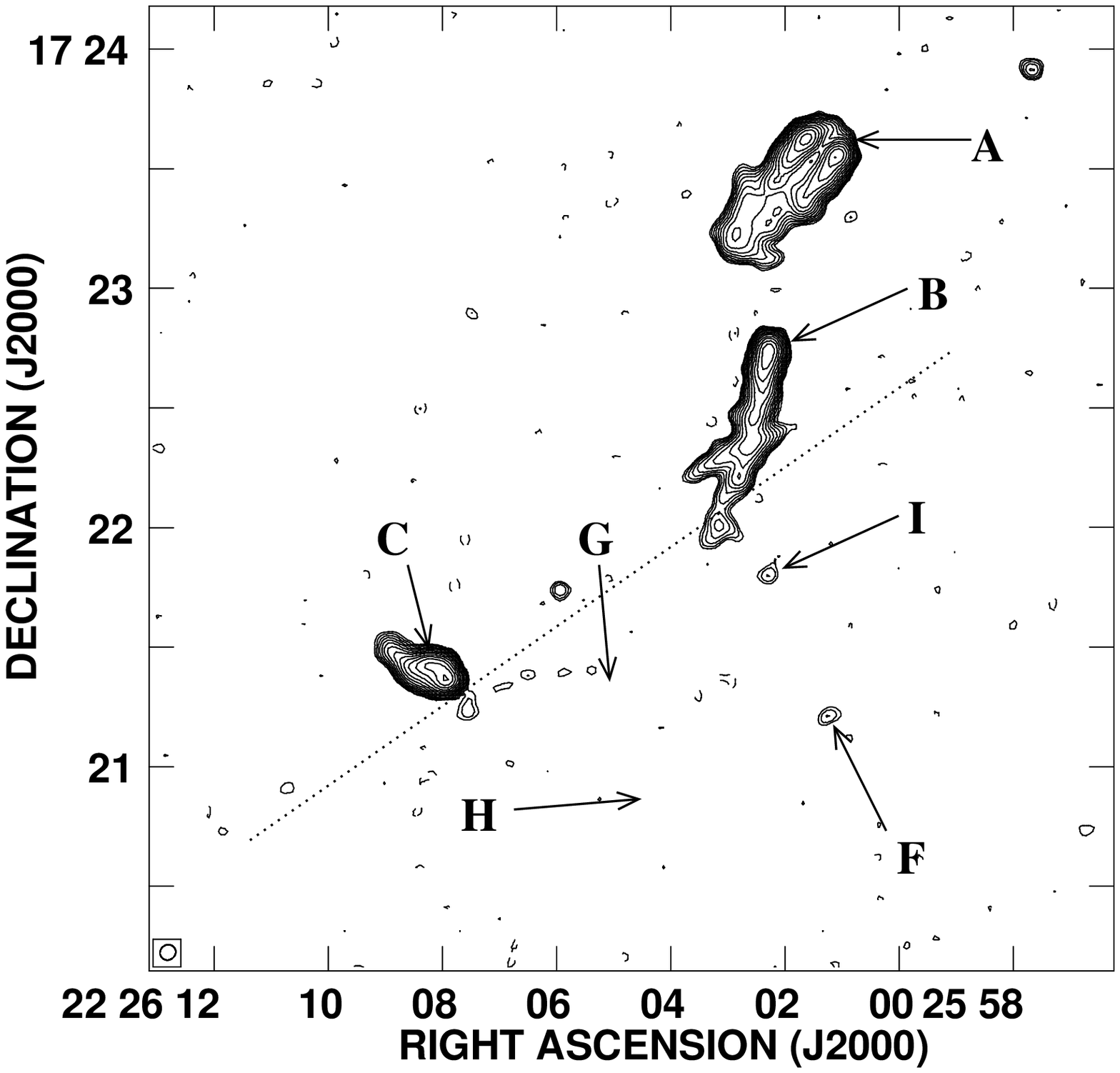}
\caption{Abell 2443 at 1425~MHz with VLA B-configuration.
Contours begin at $\pm$0.112 mJy/beam ($3\times\sigma_{rms}$) and 
increase by multiples of $\sqrt{2}$.  The peak intensity is 12.2 mJy/beam.
\label{LB.fig}}
\end{figure}

\begin{figure}
\plotone{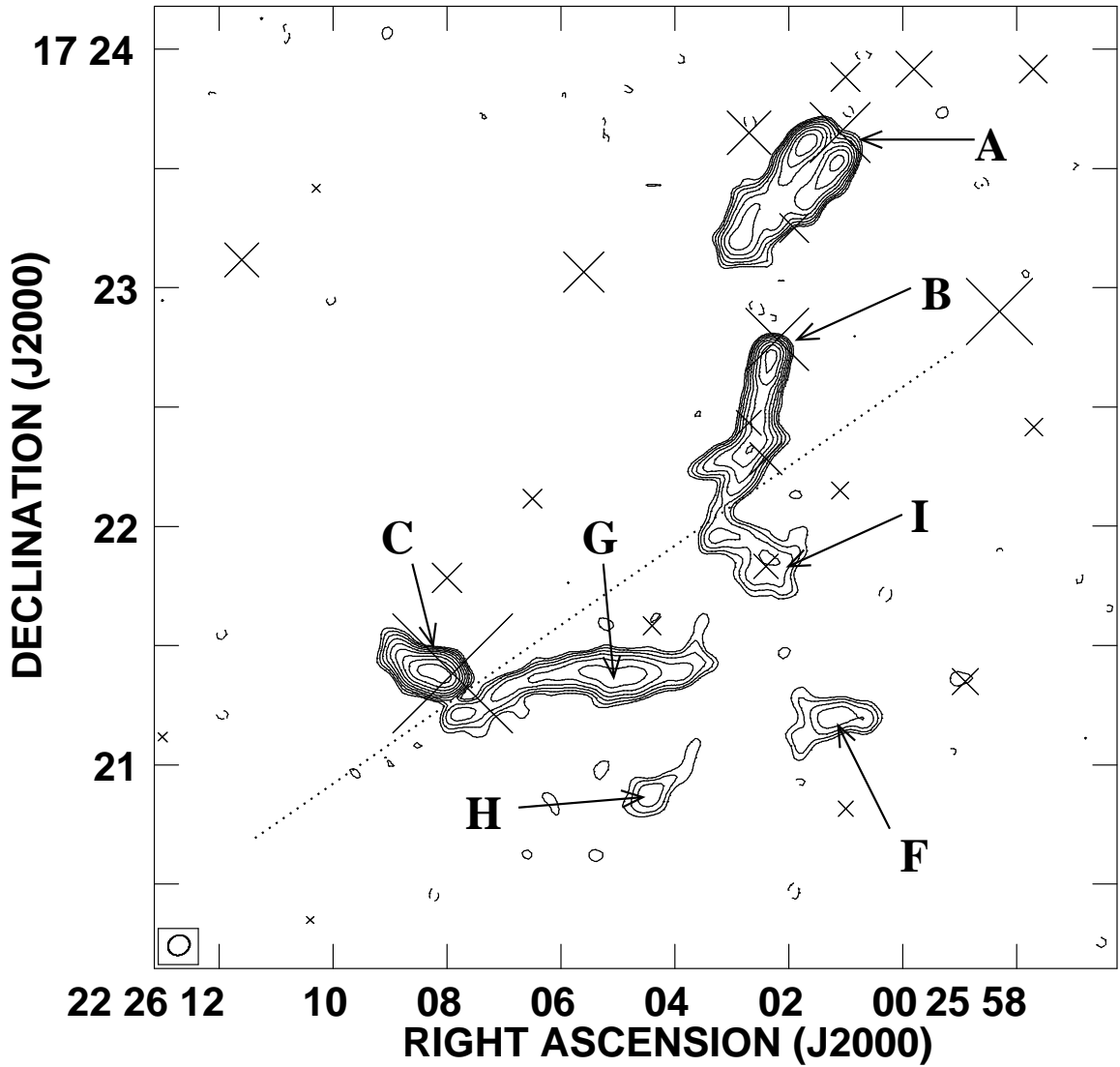}
\caption{Abell 2443 at 328.5~MHz with VLA A-configuration.
Contours begin at $\pm$1.97 mJy/beam ($3\times\sigma_{rms}$) and 
increase by multiples of $\sqrt{2}$.  The peak intensity is 38.0 mJy/beam.
The crosses indicate the locations of member galaxies and, by their sizes,
the relative K-band brightness.
\label{PA.fig}}
\end{figure}

To further examine this steep spectrum emission, higher resolution
images ($\sim$5$''$) were taken with the VLA.  At 1425~MHz this was
done in the B-configuration and at 325~MHz, this was done in the
A-configuration.  The existing 74~MHz image
was already taken in the largest configuration (A), and therefore no
higher resolution is possible at this frequency using the VLA.  These
observations served three purposes.  First, was to better determine
the morphology of this very steep spectrum emission and determine if
it is smoothly distributed or has finer scale structure.  The second
purpose was to distinguish this emission from the radio galaxies,
especially source C.  The final purpose was a better understanding of
the morphology of the radio galaxies to better distinguish their
cores, jets and lobes and their possible relationship to the steep
spectrum emission.  The observational parameters are listed in
Table~\ref{observation.tab} and the images are shown in
Figures~\ref{LB.fig} and \ref{PA.fig}.

From the higher resolution 1425~MHz image (Figure~\ref{LB.fig}), we
find that source A is apparently a narrow-angle-tail (NAT) source.
This is indicative of a significant velocity difference between the
source galaxy and the intra-cluster medium (ICM).  Source B appears as
a single head-tail source with the tail in roughly the same direction
as the tails in source A, indicating that source B may also have a
similar velocity with respect to the ICM.  Source B may also be a NAT,
but with a smaller projected angular separation between the tails than
can be resolved in this image.  Even higher resolution would be needed
to confirm this. The similar tail direction for sources A and B may be
indicative of an on-going cluster merger \citep[see e.g.][]{loken,
  gomez, burns}. We note also that the optical galaxy distribution
reveals a similar NE-SW elongation \citep{wen07}.

\begin{figure}
\plotone{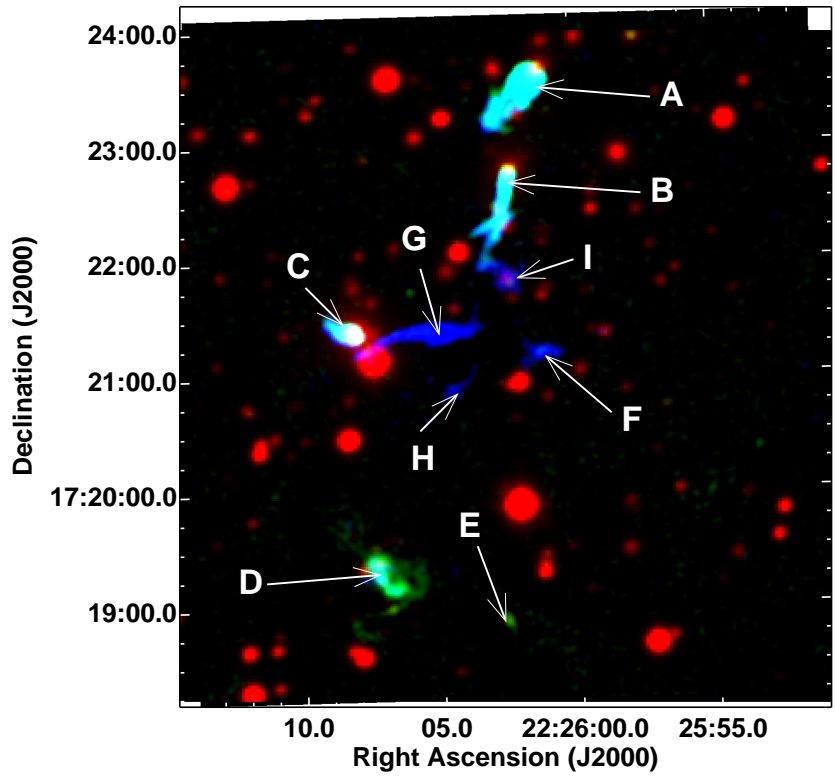}
\caption{Three-color image showing a radio/optical overlay of Abell
  2443 with R mapped to the {\it I-}band BATC image of \citet{wen07}, G
  mapped to the 1425 MHz B configuration VLA data, and B mapped to the
  328.5 MHz A configuration VLA data. The optical identifications of
  sources A, B, and C are clearly visible, as is the lack of possible
  identifications for the diffuse sources F, G, and H. Source D is
  composed of two separate radio sources, each of which have an
  associated optical source but no ID. The core of source E is
  associated with a faint galaxy as reported by \citet{slee96}.
\label{Overlay.fig}}
\end{figure}

Figure~\ref{LB.fig} also reveals a line of radio emission that crosses
the tail of source B diagonally.  This could be from another radio
galaxy, or a population of charged particles that already existed at
that location (perhaps from a past episode of activity from another
cluster member) and has been re-energized by the passing of source B.
A similar case has been seen in 3C129, another well-studied NAT source
\citep{lane02}.  This would indicate a supersonic velocity of source B
with respect to the ICM.  Source C is smaller, but has a distinct core
and a small tail fading off to the northeast.  It's velocity vector,
if any, is significantly different than that of sources A and B.  The
steep-spectrum sources F, G, and H are all either barely detected or
not detected at all.  Sources D and E are outside the region shown in
Figure~\ref{LB.fig}, and also much farther from the cluster center.
Source D appears to be comprised of two, roughly symmetric radio
galaxies, and Source E is resolved out except for the core.  These
sources are visible in Figure~\ref{Overlay.fig} but will not be
discussed further.

In the higher-resolution 325~MHz image (Figure~\ref{PA.fig}), 
sources A, B and C appear nearly 
identical to their 1425~MHz counterparts.  But there is significant radio 
emission to the south and west of the radio galaxies that did not appear
at 1425~MHz.  
The steep spectrum sources F, G, and H are now prominently 
detected, with source G appearing as a narrow but highly elongated object. 
This image shows source G to be clearly distinct from source C.  
There is also a steep spectrum source near the south portion of source
B, which we have labeled source I.  It is not clear if this emission is 
related to source B, though comparing
the 1425 and 325~MHz images shows that it is much steeper than the rest of 
the tail of source B.  If it is part of the tail, it seems like a 
discontinuous or irregular extension.  The regions of flat spectrum and 
steep spectrum emission can be divided roughly by the dotted line drawn on 
Figures~\ref{LB.fig} and \ref{PA.fig}.   

Also plotted on the higher-resolution 325~MHz image
(Figure~\ref{PA.fig}) are the locations of galaxies found to be
possible cluster members by a deep multi-band optical/IR study
\citep{wen07}.  The member galaxy positions are indicated by crosses,
where the size of the cross increases with brighter K-band magnitude
such that the cross size doubles with each magnitude of brightness.
Sources A, B and C are clearly associated with bright cluster member
galaxies located predictably at their cores.  One relatively fainter
cluster member galaxy is located within the Source I, and seems
roughly at the same location as the compact source seen in that
location in the 1425~MHz image (Figure~\ref{LB.fig}). In
Figure~\ref{Overlay.fig}, we show a three color image with the
\citet{wen07} I band image in red, the high resolution 1425 MHz VLA
data in green, and the high resolution 328.5 MHz VLA data in blue.
Inspection of this overlay shows that source I and the possible optical
counterpart are not quite aligned and therefore it is uncertain if
they are related. We further see in Figure~\ref{Overlay.fig} that
there are no optical counterparts in the remainder of the diffuse
emission and thus the steep spectrum sources located southwest of the
dotted line in Figures~\ref{LB.fig} and \ref{PA.fig} have no clear
connection to bright member galaxies.

\section{Analysis}

\subsection{Spectral Index Maps}
\label{si.maps.sec}

\begin{figure}
\plotone{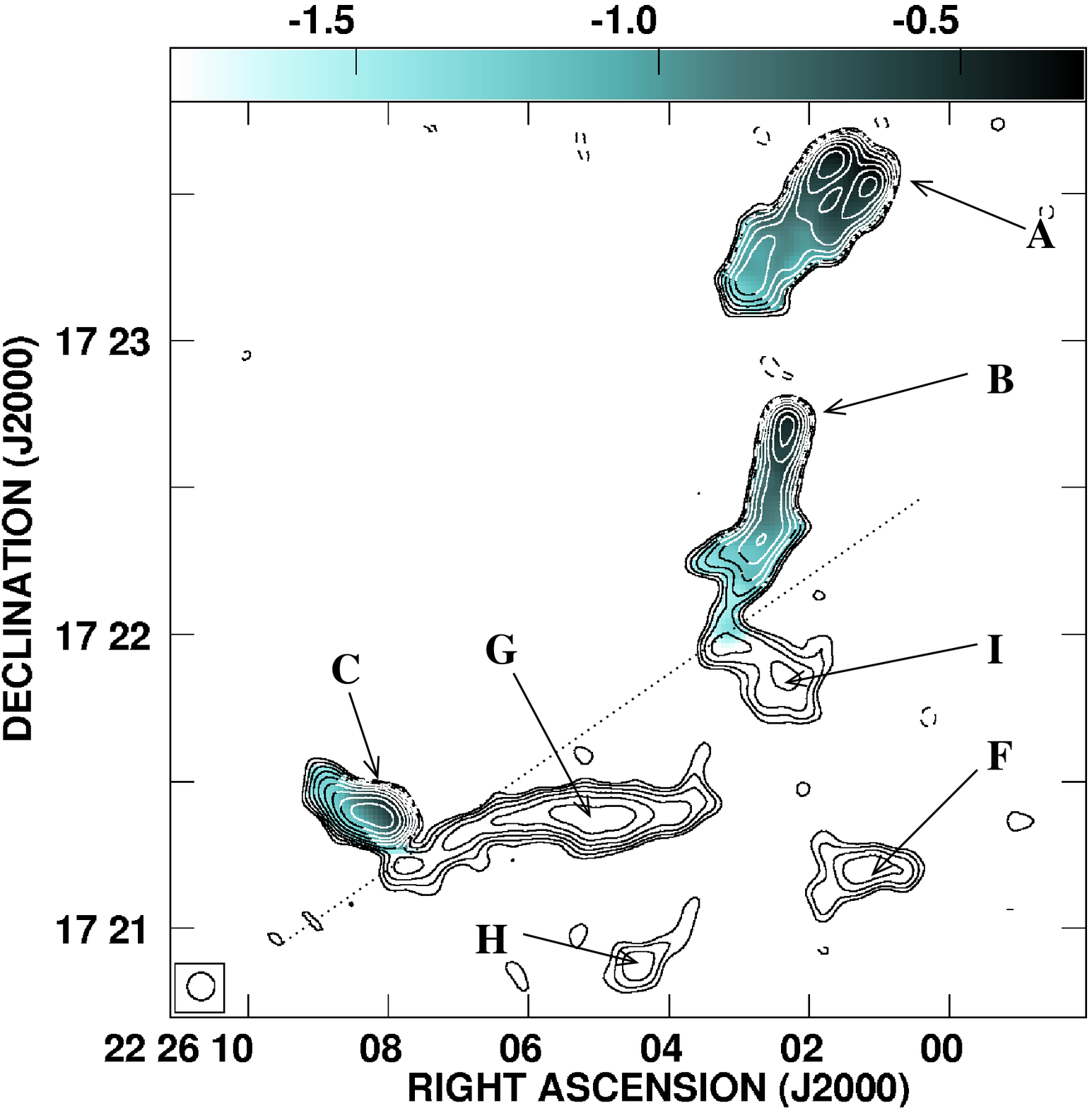}
\caption{Spectral index map (color scale) of Abell 2443 between 325~MHz 
and 1425~MHz.  The image at each frequency was convolved to a circular 5.6$''$
resolution.
Contours are for 325~MHz and begin at $\pm$2.11 mJy/beam ($3\times\sigma_{rms}$ in the convolved
325~MHz map) and 
increase by multiples of $\sqrt{2}$.  The peak intensity is 42.3 mJy/beam.
The spectral index is only shown for regions detected above  
$5\times\sigma_{rms}$ in both maps.  For regions well detected at 325~MHz that
are white, the spectral index is steeper than the lower color-scale limit of 
$\alpha_{325}^{1425} \leq -1.8$.  Because such regions are generally not 
detected or very weakly detected at 1425 MHz (Figure~\ref{LB.fig}) the 
spectral index is not well determined below this upper limit.
\label{PA.alpha.fig}}
\end{figure}

We have produced spectral index maps of Abell 2443 from our high and low 
resolution images.  The high resolution spectral index map is shown in 
Figure~\ref{PA.alpha.fig}.  This was produced by first convolving the 
1425~MHz and 325~MHz images (Figures~\ref{LB.fig} and \ref{PA.fig} 
respectively) to a common circular resolution of 5.6$''$, and only calculating
a spectral index for regions where the surface brightness was at least five
times the map noise in both images.  Most of the relic region was not 
well detected in the 1425 MHz image, and so the spectral index map does not
cover that region.  Because of the differences in sensitivity, regions that 
are detected in the 325~MHz image but not the 1425~MHz image have a spectral
index of $-1.94$ or steeper.  The radio galaxies (sources A, B and C) all 
show significant steepening along their tails, with $\alpha_{325}^{1425}$ 
ranging from roughly $-0.4$ near the core to $-1$ or steeper at the ends
of their tails.  This is again evidence that perhaps all of these sources
are some kind of NAT systems.   We also note that the line of emission that 
diagonally crosses the tail of source B has roughly the same spectral index
as the region of the tail it crosses.  This indicates that these two sources 
are of roughly the same age and probably became energized at roughly the same 
time.

\begin{figure}
\plotone{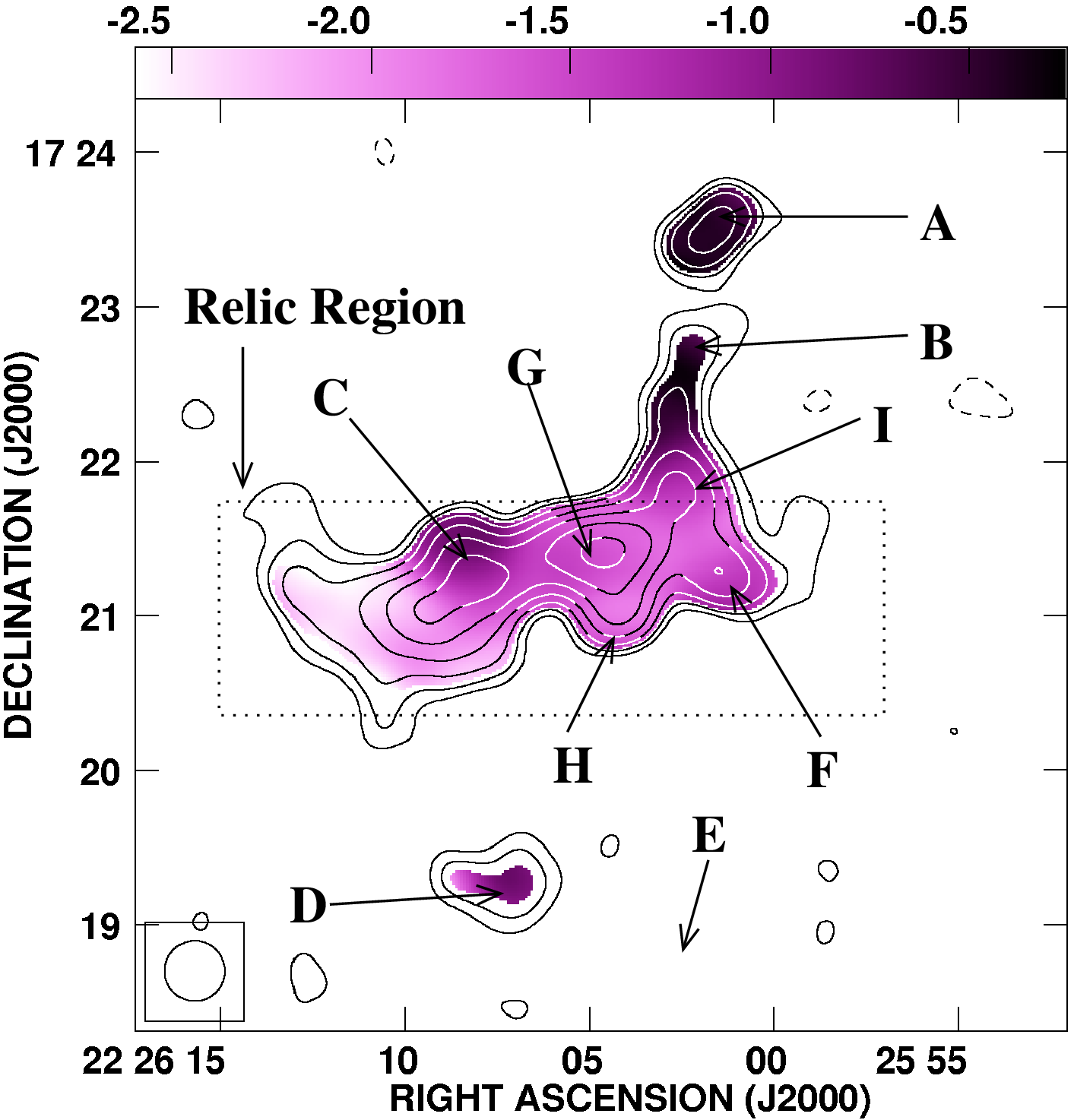}
\caption{Spectral index map (color scale) of Abell 2443 between 74~MHz 
and 325~MHz.  The image at each frequency was convolved to a circular 23.4$''$
resolution.  Contours are for 74~MHz and begin at $\pm$88.8 mJy/beam 
($3\times\sigma_{rms}$ in the convolved 74~MHz map) and 
increase by multiples of $\sqrt{2}$.  The peak intensity is 788 mJy/beam.
The spectral index is only shown for regions detected above  
$5\times\sigma_{rms}$ in both maps.  
\label{4A.alpha.fig}}
\end{figure}

We produced a lower resolution spectral index map by combining our low 
resolution maps at 74~MHz and 325~MHz, and this is shown in 
Figure~\ref{4A.alpha.fig}.   This was produced by first convolving the 
325~MHz and 74~MHz images (Figures~\ref{PB.fig} and \ref{4A.fig} 
respectively) to a common circular resolution of 23.4$''$, and only calculating
a spectral index for regions where the surface brightness was at least five
times the map noise in both images.  The spectral index map shows 
quantitatively that the emission in the relic region (with the exception 
of source C) is significantly steeper than the other emission in the cluster.
Also, the eastern end of the relic is steeper than the western end.

\subsection{Spectral Index Measurements}
\label{si.measure.sec}

Because of the stark differences in spectral indicies, one can easily 
distinguish by eye between the flat-spectrum and steep-spectrum sources.  
Here we present quantitative spectral index measurements of the various 
radio sources in Abell 2443.  

In the lower resolution images, sources A, B and C are 
difficult to distinguish completely from each other and the steep spectrum 
emission.  This is much easier in the higher resolution images, and so 
we use these maps to measure the flux densities of these sources at 
325~MHz and 1425~MHz.  Because these are radio galaxies, rather than relic 
emission, we do not expect significant diffuse emission beyond the $\sim 5''$ 
scale of the higher resolution images, and therefore it is unlikely that 
measuring the flux density in the higher resolution images misses 
significant flux density.
Because we have no high-resolution map at 74~MHz, we used the low-resolution 
74~MHz image (Figure~\ref{4A.fig}) to measure the flux densities of sources
A, B and C by summing the flux density within small boxes drawn around their
known location from the higher resolution images.  We note that this method 
is only an estimate of their flux densities, because this method, especially 
for source C, was more susceptible to contamination from nearby sources.  
However, the 74~MHz flux density of source C is only about 5\% that of the 
relic emission, so even a significant error here would not significantly 
contaminate the measurement of the relic flux density which follows.

Next we combine all the steep spectrum emission into a single source that 
we simply call the ``relic'' source.  The relic encompasses not just 
sources F, G and H, but also all the steep spectrum emission seen in the 
rectangular relic region indicated in Figures \ref{LC.fig}, \ref{PB.fig} 
and \ref{4A.fig}.  The flux
density of the relic was determined using the low-resolution maps (to avoid 
missing diffuse emission) by summing the flux density in the relic region 
and subtracting the flux density of the flat-spectrum source C.  

\begin{figure}
\plotone{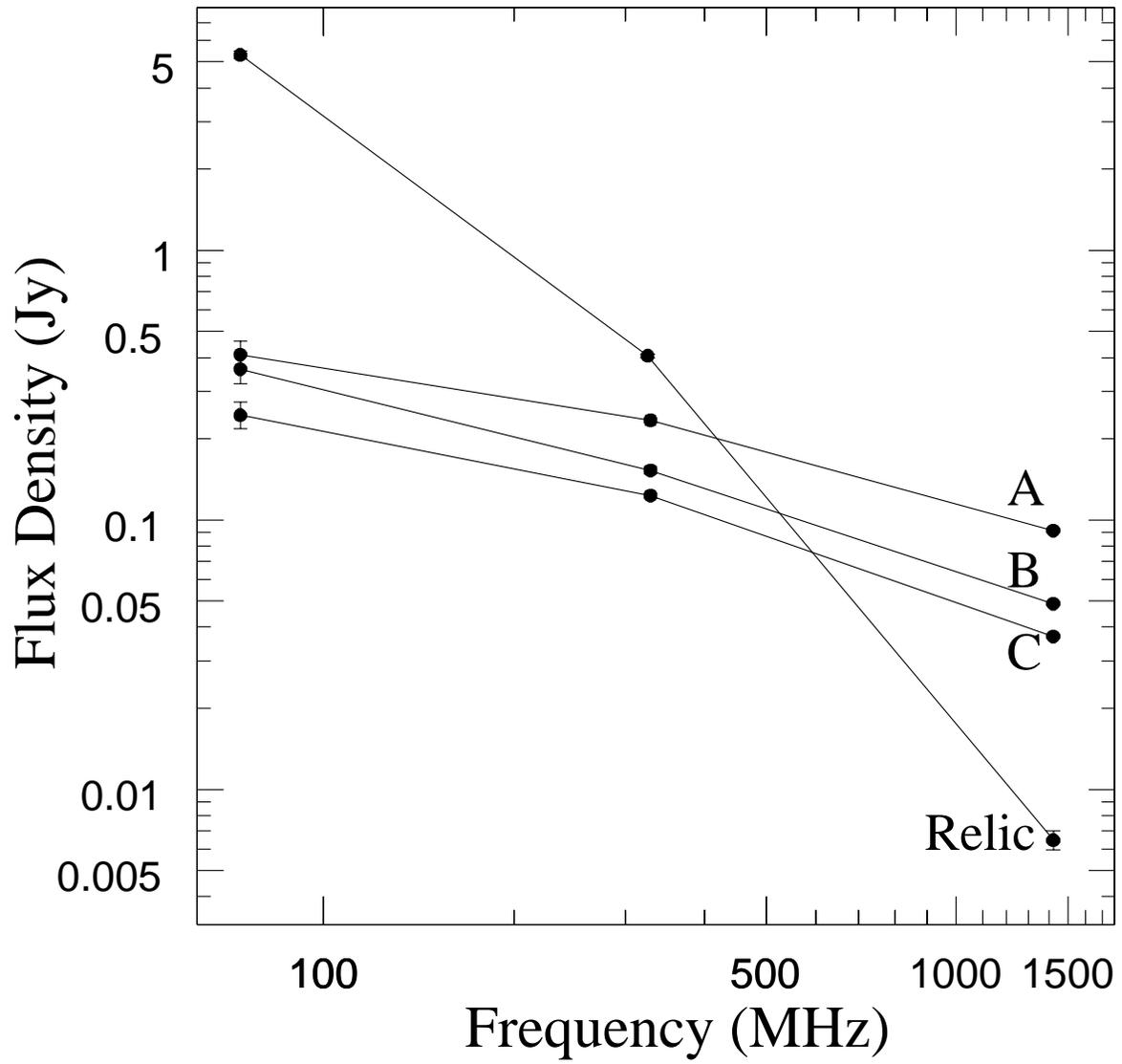}
\caption{Flux densities as a function of frequency for the major components 
of galaxy cluster Abell 2443.
\label{flux.fig}}
\end{figure}

The flux densities are plotted in Figure~\ref{flux.fig} and listed in
Table~\ref{flux.tab} along with the resulting spectral index
measurements.  For the radio galaxies (sources A, B and C), we find
spectral indices between 325~MHz and 1425~MHz ($\alpha_{325}^{1425}$)
ranging from $-0.637$ to $-0.813$, which is typical for radio
galaxies.  Between 74~MHz and 325~MHz, the spectra flattens a bit with
$\alpha_{74}^{325}$ ranging from $-0.379$ to $-0.584$.  This
low-frequency flattening is also commonly seen in radio galaxies
\citep{kk66}.  The relic emission is dramatically steeper with
$\alpha_{325}^{1425} = -2.797$.  This is steeper than the typical
relic found in \citet{feretti05}, yet well within the range of the
``extreme'' relics found by \citet{slee01}, which have $-4.4 < \alpha
< 2.1$, measured at frequencies near 1400~MHz.  In the lower frequency
interval, the spectrum flattens for the relic also, but is still
extremely steep at $\alpha_{74}^{325} = -1.737$. This curved spectral
signature is characteristic of the radio phoenix class of relics
\citep{kempner04} which have experienced synchrotron and
inverse-Compton radiation losses prior to re-energization by the
passing shock \citep{ensslin01}. Such curved spectra are difficult to
produce for radio gischt, which are the result of shock acceleration
from the thermal electron reservoir \citep{hoeft07}

\subsection{Small-Scale Structure in the Relic}

The best high-resolution image of the relic is at 325~MHz 
(Figure~\ref{PA.fig}).  This is because the relic spectral index is too 
steep for it to be well-detected at 1425~MHz.  At 74~MHz, the VLA is not 
capable of such high resolution.  At 325~MHz, we see significant 
smaller-scale structure including isolated peaks and a roughly 1-arcminute 
arc of emission at the location of source G.  The 1-arcminute extent
corresponds to a projected physical size of roughly 120 kpc.  The sum of the 
steep spectrum 
emission is 205 mJy, which is roughly half of what was detected in the
lower-resolution 325~MHz image (Figure~\ref{PB.fig}).  This indicates that 
about half of the emission is resolved out in the 
higher resolution image.  We note that the extremely steep spectrum, the 
physical size and the small scale structure in this relic makes this source
very similar to previous extreme relics as examined by \citet{slee01}.

In addition to showing that the relic emission is not smoothly
distributed, the small-scale structure is possibly indicative of the
origins of the relic.  In particular, sources G and I appear to be
extensions of sources C and B, respectively.  However, in both cases,
the extension seems discontinuous.  Also, in both cases, this
discontinuity occurs at the dividing line between the flat and steep
spectrum emission in the cluster.  Finally, in both cases, the
discontinuity is the result of the emission on the steep-spectrum side
seemingly being shifted to the south-east along this dividing line. An
alternative scenario for sources G and I is that they are re-energized
sources from adiabatic compression of the tail of source B which
appears to be moving rapidly with respect to the ICM. The meaning of
these observations is not clear, and it is likely that more extensive
observations in both radio and X-ray will be necessary to explain the
origin of these features.

\subsection{X-ray Data}

The best X-ray data available on this cluster comes from the ROSAT All
Sky Survey \citep[RASS;][]{voges99}.  We show this map convolved to
120$''$ overlaid on our low-resolution 325~MHz image of the cluster.
This shows the center of the X-ray emission appears roughly at the
location of source B, and that the relic emission is offset from the
cluster center, which is typical of both the phoenix and gischt
classes of radio relics.  At its closest point, the relic emission is
separated from the center of the cluster X-ray emission by a projected
distance of 160 kpc.

The sensitivity and resolution of this X-ray image are not 
sufficient to indicate either merger activity or shock fronts in the ICM.  
In order to further investigate the nature of the radio relic and its relation
to the cluster, it is extremely important to have a more accurate image of 
the X-ray emission.  

\subsection{Evidence for a Merger}
There is no conclusive evidence that Abell 2443 is experiencing or has recently
experienced a merger or collision.  The available X-ray data are not 
sufficient to show shocks or other asymmetries in the ICM.  A multi-band 
optical/IR study of the galaxies in this region shows some evidence that 
another cluster, ZwCl 2224.2+1651, might be falling towards Abell 2443 from 
the southeast \citep{wen07}.   But the projected distance to 
ZwCl 2224.2+1651 is 2.5~Mpc, which makes it unlikely that this in-fall could 
have caused the relic we observe.  

However, other evidence strongly suggests a major ongoing merger.
First is the apparent high velocity difference between the ICM and the
galaxies producing radio sources A and B.  Second, the fact that there
are 3 large and active galaxies (radio sources A, B and C) is also
suggestive of recent or ongoing merger activity since typical clusters
are only expected to have 0 to 2 head-tail galaxies \citep{lo95}.
 
\section{Conclusion}
We have presented deep radio images at 1425, 325 and 74~MHz of radio
emission in the galaxy cluster Abell 2443.  These maps reveal a region
of emission that is (1) diffuse, (2) extremely steep spectrum (3) on
the cluster periphery (4) of irregular morphology and (5) not directly
associated with any of the galaxies within the cluster.  We conclude
that this emission is most likely an ultra-steep spectrum cluster
radio relic.  The high-resolution radio maps show morphology in the
relic and member radio galaxies consistent with a recent or ongoing
cluster merger. The curved spectral shape of the relic combined with
the relatively small scale and compact structure suggest that it is
likely a member of the radio phoenix class outlined in
\citet{kempner04}. More accurate X-ray data are required to confirm
the cluster dynamical state and establish the relation between the
observed radio features and the merger-related activity within the
ICM.

\section{Acknowledgments}

We thank Zhong-Lue Wen for providing the {\it I-}band BATC image of
Abell 2443 as well as the cluster member catalogue. We thank Torsten
Ensslin for helpful discussions. Basic research in radio astronomy at
the Naval Research Laboratory is supported by 6.1 Base funding.  The
National Radio Astronomy Observatory is a facility of The National
Science Foundation operated under cooperative agreement by Associated
Universities, Inc.  We have made use of the ROSAT Data Archive of the
Max-Planck-Institut f\"{u}r extraterrestrische Physik (MPE) at
Garching, Germany. We thank the referee for helpful comments and
suggestions which improved this paper.

\vfill
\eject

\vfill
\eject

\vfill
\eject

\vfill
\eject

\vfill
\eject

\vfill
\eject

\vfill
\eject

\vfill
\eject

\vfill
\eject
\vfill
\eject

\vfill
\eject

\begin{figure}
\plotone{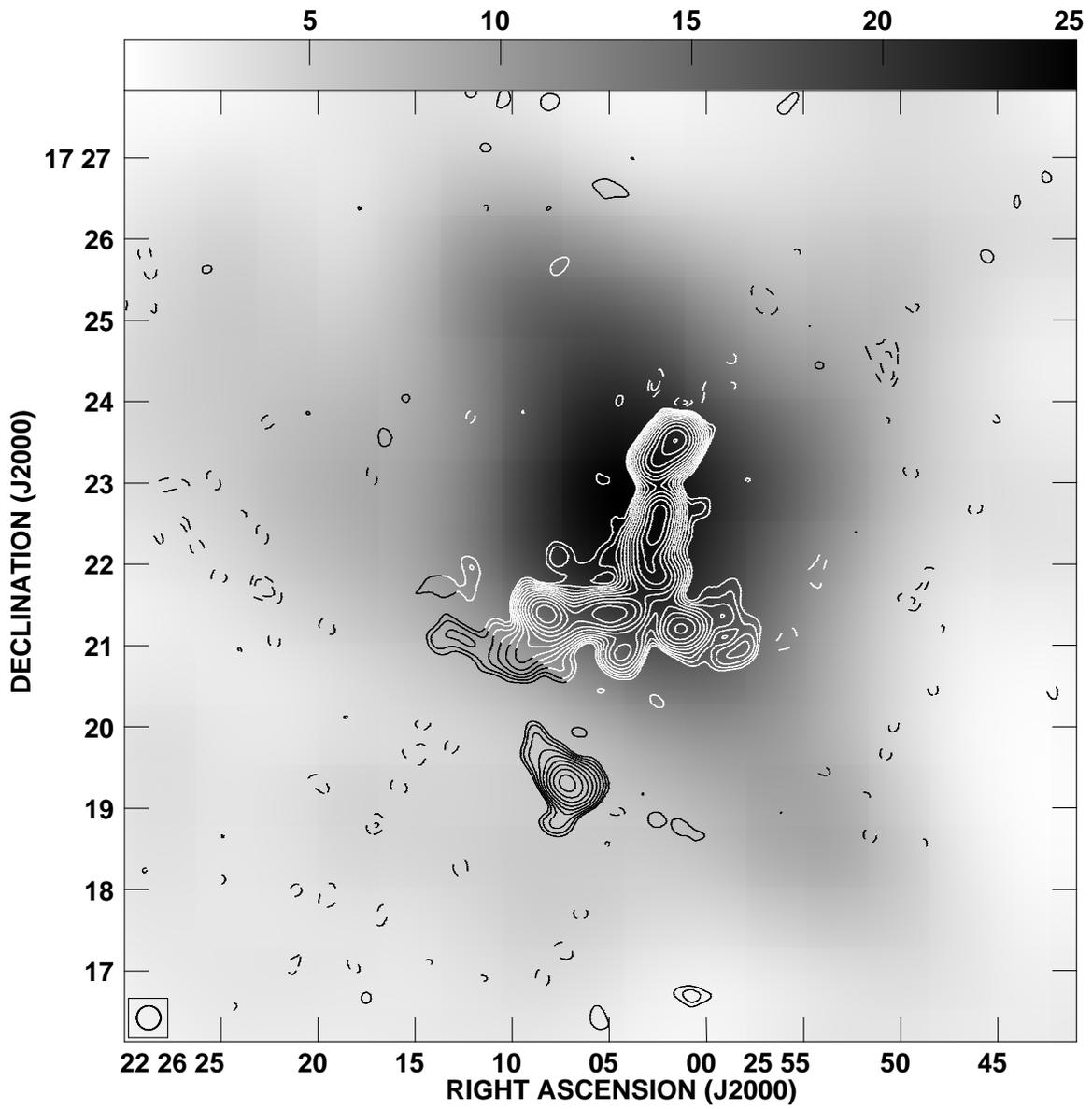}
\caption{Abell 2443 at 325~MHz (contours) overlaid on ROSAT All Sky Survey 
\citep[RASS;][]{voges99} image convolved to a resolution of 120$''$.  
\label{pb.rass.fig}}
\end{figure}
\vfill
\eject

\input{tab1.tex}

\input{tab2.tex}

\end{document}

%% file: tab1.tex
\begin{deluxetable}{lrrrrrr}[tb]
\tablecaption{VLA Observations of Abell 2443
\label{observation.tab}}
\tablehead{
\colhead{Date} & 
\colhead{Code} & 
\colhead{$\nu$} & 
\colhead{c} & 
\colhead{TOS} &
\colhead{$\sigma_{rms}$} &
\colhead{Resolution}}
\startdata
\multicolumn{7}{c}{Low Resolution} \\
\hline
2005-07-03 & AC786 & 1425 & C & 115 & 0.0493 & $13.4\times12.2$,-24\\
2006-07-24 & AC822 & 325 & B & 260 & 0.803 & $17.5\times15.9$,-67 \\
2007-09-01 & AC882 & 73.8 & A & 344 & 29.4 & $23.4\times22.9$,-21 \\
\hline
\\
\multicolumn{7}{c}{High Resolution} \\
\hline
2006-07-24 & AC822 & 1425 & B & 140 & 0.0372 & $4.00\times3.74$,-54 \\
2007-09-01 & AC882 & 325 & A & 344 & 0.656 & $5.56\times4.90$,-62 \\
\enddata
\tablecomments{For each observation: $\nu$, is the central frequency in MHz, $c$ refers to the configuration of the VLA, TOS is the time on source in minutes, $\sigma_{rms}$ is the map noise in mJy/beam, and Resolution is the synthesized beam dimensions in arcseconds with the position angle in degrees. }
\end{deluxetable}

%% file: tab2.tex
\begin{deluxetable}{lrrrrr}[tb]
\tablecaption{VLA Observations of Abell 2443
\label{flux.tab}}
\tablehead{
\colhead{Source} & 
\colhead{$S_{74}$} & 
\colhead{$S_{325}$} & 
\colhead{$S_{1425}$} & 
\colhead{$\alpha_{74}^{325}$} & 
\colhead{$\alpha_{325}^{1425}$}}
\startdata
A & $410$ & $234$ & $91.3$ & $-0.38$ & $-0.64$ \\
 & $\pm51$ & $\pm5.0$ & $\pm0.38$ & $\pm0.05$ & $\pm0.01$ \\
B & $363$ & $153$ & $48.9$ & $-0.58$ & $-0.77$ \\
  & $\pm49$ & $\pm4.9$ & $\pm0.37$ & $\pm0.06$ & $\pm0.01$ \\
C & $244$ & $123$ & $37.0$ & $-0.46$ & $-0.81$ \\
  & $\pm29$ & $\pm2.3$ & $\pm0.13$ & $\pm0.05$ & $\pm0.01$ \\
Relic & $5310$ & $406$ & $6.5$ & $-1.74$ & $-2.80$ \\
 & $\pm175$ & $\pm69$ & $\pm0.54$ & $\pm0.07$ & $\pm0.09$ \\
\enddata
\tablecomments{All flux densities are given in mJy.}
\end{deluxetable}